\RequirePackage{amsmath}
\RequirePackage{url}
\documentclass[11pt,letterpaper]{IEEEtran}
\usepackage[numbers,sort]{natbib}

\usepackage{xspace}

\usepackage[inline]{enumitem}

\usepackage{xcolor}
\definecolor{listbackgroundcolorlight}{rgb}{0.91,0.92,0.94}
\definecolor{mpsNavy}{rgb}{0.01,0.01,0.5}%
\definecolor{mpsGray}{rgb}{0.85,0.84,0.84}%
\definecolor{mpsGreen}{rgb}{0.01,0.4,0.01}%
\definecolor{mpsRed}{rgb}{0.70,0.01,0.01}%

\usepackage{listings}
\usepackage{tikz}
\usepackage{standalone}
\usetikzlibrary{calc}
\usetikzlibrary{fit}
\usetikzlibrary{positioning}

\usepackage{pgfplots}
\pgfplotsset{compat=1.17}

\newcommand{\fsl}{\textsl}

\newcommand{\fsf}[1]{{\footnotesize{\textsf{#1}}}}

\newcommand{\pname}[1]{\fsl{#1}}
\newcommand{\sname}[1]{\fsl{#1}}
\newcommand{\ename}[1]{\fsl{#1}}
\newcommand{\rname}[1]{\textsc{#1}}
\newcommand{\paraname}[1]{\fsf{#1}}

\usepackage[normalem]{ulem}

\newcounter{mpscount}
\newcounter{akccount}
\newcounter{shccount}

\newcommand{\review}[1]{}
\newcommand{\response}[1]{}

\newcommand{\speciallabelsize}{\normalsize\rm}
\newenvironment{exfacts}[1]{\begin{list}{{\speciallabelsize
	\theenumi.}}{\usecounter{enumi}
      \settowidth{\labelwidth}{{\speciallabelsize #199}}
      \setlength{\leftmargin}{\labelwidth}
      \addtolength{\leftmargin}{0.02\labelsep}}}{\end{list}}

\newcounter{norms} 
\newcommand{\bnorm}{\begin{exfacts}{Norm}\setcounter{enumi}{\value{norms}}\renewcommand{\theenumi}{$\mathbf{N_\arabic{enumi}}$}}
\newcommand{\enorm}{\setcounter{norms}{\value{enumi}}\renewcommand{\theenumi}{
    \arabic{enumi}.}\end{exfacts}}

\title{Hercule: Representing and Reasoning about Norms as a Foundation for Declarative Contracts over Blockchain}

\author{\IEEEauthorblockN{Samuel H.~Christie~V}, \and
\IEEEauthorblockN{Amit K.~Chopra}, \and
\IEEEauthorblockN{Munindar P.~Singh}}

\begin{document}


\maketitle

\begin{abstract}
Current blockchain approaches for business contracts are based on smart contracts, namely, software programs placed on a blockchain that are automatically executed to realize a contract. However, smart contracts lack flexibility and interfere with the autonomy of the parties concerned.

We propose Hercule, an approach for declaratively specifying  blockchain applications in a manner that reflects business contracts.  Hercule represents a contract via regulatory norms that capture the involved parties' expectations of one another. It computes the states of norms (hence, of contracts) from events in the blockchain.  Hercule's novelty and significance lie in that it operationalizes declarative contracts over semistructured databases, the underlying representation for practical blockchain such as Hyperledger Fabric and Ethereum.  Specifically, it exploits the map-reduce capabilities of such stores to compute norm states.

We demonstrate that our implementation over Hyperledger Fabric can process thousands of events per second, sufficient for many applications.
\end{abstract}


\section{Introduction}
\label{sec:introduction}

A contract conceptually underpins any application that involves two or more autonomous parties. Blockchain, by providing shared state and event ordering across trust boundaries, can enable shared and automated interpretation and adjudication of contracts.

Today's blockchains support \emph{smart contracts}, software programs meant to automate contracts \cite{Szabo-97:smart}. However, smart contracts prove unwieldy for assurance; in combination with the
immutability of blockchain, they can yield disastrous outcomes, as the DAO incident \citep{Mehar2019:DAO-Attack} illustrates.
In conceptual terms, smart contracts obstruct the autonomy and flexibility of the parties to a business transaction,
which is simply unacceptable in real-world applications
\citep{Computer-20:Violable}.
Ricardian contracts (\url{https://iang.org/papers/ricardian_contract.html}) tie a natural language description with a computational representation, possibly a smart contract.
They face the fundamental problem of confusion about which version is correct: the one a human can read or the one the blockchain executes.
Instead, we develop a declarative model to respect autonomy, improve readability, reduce bugs, and avoid computing the effects of each transaction on every validator node.

We build on recent work on a declarative representation for contracts based on \emph{regulative norms} that provides high-level abstractions with a precise semantics \citep{Computer-20:Violable}.
``Norms'' here are not mere descriptions of social behavior but carry prescriptive force and serve as elements of legal contracts \citep{Von-Wright-63:Norm,Ailaw-99}.
Norms have been formalized and mapped to a relational information model of events \cite{AAMAS-16:Custard}.
However, the relational approach has crucial limitations: (1) it requires a fixed structure of tables and columns and (2) it is often unavailable.
For example, Hyperledger Fabric, a leading blockchain architecture, provides only a LevelDB based key-value store, and a CouchDB-based document store.


Therefore, to realize declarative contracts on blockchain platforms, it is important to show how a logic-based contract language can be implemented over a semistructured information model.  We consider document stores wherein documents generally lack a schema, do not follow a predefined structure, and do not support the indexes required for efficient join queries common for relational settings.  Instead, document stores process queries using map and reduce operations for data parallelism.

\subsection{Contributions}
We contribute Hercule, an approach to realize contracts as norms over event histories that can be securely shared between multiple agents using a blockchain. We show how to automatically generate map-reduce queries from norms.

We empirically evaluate our implementation and show that it is efficient enough to be used in any situation where Hyperledger Fabric is practical, while providing the requisite expressiveness and flexibility.

\section{Contracts and Smart Contracts}
\label{sec:Smart}

A \emph{contract} is a document describing a legal relationship between multiple parties stating what each party may expect from the others under what conditions.

A \emph{smart contract} is a programmatic description of a contract along with an architecture for automatic verification and enforcement. For example, a vending machine that automatically provides a product to anyone who pays the requisite amount is a smart contract \cite{Szabo-97:smart}.

Bitcoin, the first blockchain \citep{Bitcoin}, represents financial transactions as simple, limited programs that verify the claims made against them.
Subsequent blockchain platforms, such as Ethereum and Hyperledger Fabric, extend the concept of smart contracts to arbitrary executable programs, often called \emph{distributed applications} or Dapps.

However, specifying contracts as arbitrary programs has major drawbacks.
First, since the program is the source of truth regarding the meaning of a smart contract, there is no recourse against unintended behaviors, as the infamous DAO incident highlighted \citep{Computer-20:Violable}.
Second, a program omits the portions of a business relationship that cannot be automated, limiting the participants' ability to exercise their autonomy. 
In general, a party to a contract would exercise its discretion in acting, including deciding what terms to violate and penalties to risk.

Automation is infeasible wherever human insight is needed.
For example, even if sharing a patient's health information is normally prohibited, a physician might share the information with a specialist during a medical emergency; an automated system would be unable to violate the prohibition.
An auditor can decide afterward whether the physician was justified in violating the prohibition.
Automation is also infeasible if external resources are involved.
For example, a hospital may borrow supplies from another hospital promising to return them on request.
To guarantee automatic return, a Dapp must control the supplies and not let them be used, which would defeat the purpose of borrowing.

\section{Compacts and Regulatory Norms}
\label{sec:Sidebar}

Hercule seeks to capture the essence of real-life contracts by specifying the legal relationships between the parties without curtailing their autonomy. We adopt the name \emph{compact} for our formal notion to differentiate it from natural language contracts.

We summarize established terminology and semantics on
norms \cite{AAMAS-16:Custard}.  A norm (instance) is a directed expectation between two agents, the \emph{expector} and \emph{expectee}.  A norm is generally conditional, featuring an \emph{antecedent} and a \emph{consequent}, both events (possibly complex: logical expressions over simpler events). Three major norm types, \emph{commitment}, \emph{prohibition}, and \emph{authorization}, are adequate for illustrating Hercule's language and reasoning.  

To illustrate these norm types, we adopt a scenario from healthcare as our running example.  Healthcare providers (HCPs) possess electronic health record (EHR) data about their patients that could be useful in research, but privacy regulations prevent them from sharing that data to outside organizations, unless specifically authorized by the patient. We motivate four main requirements on HCPs, which we formalize via norms: storing data, destroying data, enabling legitimate access, preserving confidentiality.

In a \textbf{commitment} norm, the expectee (\emph{debtor}) commits to the expector (\emph{creditor}) that if the antecedent occurs, then the consequent will occur.  If the antecedent occurs but the consequent doesn't, then the commitment is violated; if the consequent occurs, it is satisfied. \ref{norm:StoreData} and \ref{norm:DestroyData} below specify commitments in our EHR scenario compact.

\bnorm
\item\label{norm:StoreData}
(StoreData) An HCP commits to store a patient's data (consequent) after they visit (antecedent).
The HCP is the debtor and the patient is the creditor; if a patient visits but the HCP does not store their data, the HCP violates the commitment.

\item \label{norm:DestroyData} (DestroyData)
The HCP commits to destroying a patient's data (which must not be stored directly on the blockchain!) upon request by the patient.
\enorm

In a \textbf{prohibition} norm, the expectee is prohibited by the expector from bringing about the consequent if the antecedent holds.
If the antecedent and the consequent both occur, then the prohibition is violated; if the antecedent occurs but not the consequent, it is satisfied.
\ref{norm:Access} in our EHR compact is a prohibition.
\bnorm
\item \label{norm:Access}
(Access)
An HCP is prohibited from sharing patient data with an agent (consequent) unless the patient has granted access (antecedent).
\enorm

In an \textbf{authorization} norm, the expectee authorizes the expector for the consequent when the antecedent occurs.
If the antecedent occurs, then if the consequent cannot occur (e.g., due to incorrect access control settings), then the authorization is violated, else it is satisfied.
\ref{norm:Confidentiality} in our EHR compact is an authorization.
\bnorm
\item \label{norm:Confidentiality}
(Confidentiality)
A hospital may authorize a family member for access to a patient's health records (consequent) if the patient submits a release form (antecedent).
\enorm

Any norm is instantiated (created) as specified events occur, but expires if its antecedent doesn't occur. For example, when the patient sends a specific directive identifying the data item and recipient, the authorization is instantiated for that patient, that item, and that recipient.

\section{Declarative Contracts Using Norms}
\label{sec:Declarative}

We adopt a simplified notion of the components and lifecycle of a compact \cite{Vazquez-JAAMAS-05}:

\begin{description}
\item [Provisions] or the clauses describing what is expected of or offered to each party.
\item [Operational model] or how participants may move in reference to a compact and the outcomes of their moves.
\item [Enforcement] or how violations are handled.
\end{description}

Hercule implements compacts specified using norms to represent the provisions of a compact, which yield flexible operationalization and enforcement.
Although Hercule does not provide tools for specifying the operational model or enforcement (outside of defining additional provisions), these components can be deployed as part of the implementation.
This approach to implementing operational constraints is similar to defining the standard procedures of a legal jurisdiction in which a compact is interpreted; not every compact needs to specify the procedures used in the courtroom for adjudication, because they are assumed from the legal context.
From this perspective, Hercule is a framework for implementing a jurisdiction or legal context in which compacts can be formed, leaving the exact operational and enforcement models largely out of scope for the compacts themselves.

We propose using regulative norms as a model for higher-level declarative specifications that capture intentions and relationships, instead of low-level procedures.
Our model shares an intuition with Bitcoin, in which events are recorded to the blockchain but the interpretations are made by the clients.
Specifically, Bitcoin does not use scripts to update balances stored on the blockchain; indeed it does not store user balances at all.
Instead, it records individual transactions, and users compute their total balances by aggregating the relevant transactions.
In the same way, we use the blockchain as a history of events that can be validated and preserved in themselves, with the interpretation supplied externally by a client.
External interpretations are acceptable, because ultimately enforcement will be carried out by some external authority (which may be an automated service); this is the same as traditional contracts, where each party may have their own opinions about whether a violation has occurred, but the ultimate decision belongs to an arbiter.

Unlike most smart contract systems, which require all nodes to redundantly compute results during transaction processing, in Hercule only those clients that care about a given compact state are required to compute it.
Also, Hercule separates the enactment of a compact from its enforcement, enabling flexible social enforcement methods and updates to the compact that reinterpret past events without forking the history.

Hercule builds on a concept of histories of events, related by logical propositions.
Using this simple foundation, we build up to a normative framework with explicit support for commitments, authorizations, and prohibitions as well as the flexibility to define custom norm types.

\subsection{Architecture}

\begin{figure}[htb]
\centering
\begin{adjustbox}{max size={\columnwidth}{!}}
\includegraphics{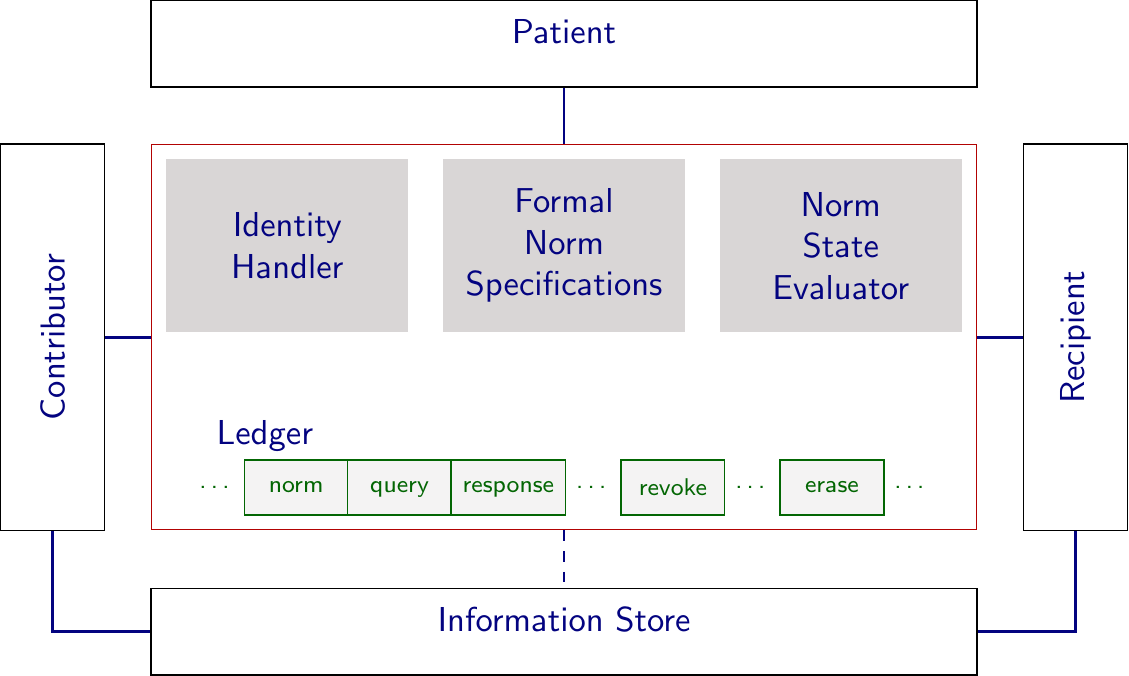}
\end{adjustbox}
\caption{A high-level schematic description of a potential Consent
  Management System built on Hercule and how it interfaces with the
  main stakeholders and with external information stores.  Here, a
  contributor is a patient's physician or laboratory or clinical trial
  firm; a recipient is a researcher.  Hercule provides norm
  specifications and a norm state evaluator overlaid on a ledger that
  captures a consensus view of events.  The identity handler maps a
  patient identity to identifiers in external information stores,
  and is not emphasized in this paper.}
\label{fig:consent-architecture}
\end{figure}

Whereas Hercule can be applied in multiple settings, Figure~\ref{fig:consent-architecture} illustrates the architecture of a consent management system built with Hercule and deployed on top of a distributed ledger such as Hyperledger Fabric.

\subsection{Events}
Events are immutable descriptions of information observed by an agent.
Hercule represents events as JSON documents.

\begin{lstlisting}[label={list:event-instance},caption={Example event object.}]
"Store": {
  "$\$$by": "H",
  "item": "Diagnosis",
  "$\$$time": 3
}
\end{lstlisting}

Listing~\ref{list:event-instance} shows a single instance of the \ename{Store} event.
Importantly, each event is recorded by a specific agent at a specific time, represented here by the \paraname{\$by} and \paraname{\$time} attributes.
For simplicity, we represent time as an integer.

\subsection{Histories}
\label{sec:Histories}
A history is a record of the information produced in an interaction between multiple parties.
In Hercule, a history is represented as a single JSON document with a unique identifier in which attributes and values are only added and never modified.
A single ledger may record multiple histories that are mutually unrelated.
All events and norms are interpreted relative to a given history.

New information is added to a history through a transaction on the ledger when an event occurs.
When updating a history, both the new event and copies of the new information it produces are added to the history as attributes.
Although binding attributes on the history itself adds some overhead, doing so generalizes information away from the events that produce it, so that compact provisions can be written to depend on the information instead of the specific events.

In Hercule, a history is represented as a single JSON document.
Listing~\ref{list:enactment-instance} shows an example history, building on the previous \ename{Store} event.
The history has a unique ID, and comprises several events each represented by a subdocument.
Each subdocument has attributes and values corresponding to information parameters bound by one of the events.

\begin{lstlisting}[label={list:enactment-instance},caption={An example history document.}]
{
  "_id": "7c4f054dc8740698c93a9452e856cc87",
  "patient": "P",
  "physician": "D",
  "hospital": "H",
  "date": "2018-11-16",
  "item": "Diagnosis",
  "Visit": {
    "$\$$by": "P",
    "date": "2018-11-16",
    "$\$$time": 1,
  },
  "Record": {
    "$\$$by": "D",
    "patient": "P",
    "item": "Diagnosis"
    "$\$$time": 2,
  },
  "Store": {
    "$\$$by": "H",
    "item": "Diagnosis"
    "$\$$time": 3,
  }
}
\end{lstlisting}

The document representation matches the map-reduce querying facility of the CouchDB database underlying a Hyperledger Fabric ledger.
Map and reduce operate only on individual documents in isolation, so any information that needs to be correlated for computing a query result must be stored together.

\subsection{Norms}
\label{sec:Norms}
Norms are specified as a set of \emph{states}, which are logical formulas over events, usually involving multiple parties.
A norm is \emph{instantiated} in a history if any of its states are satisfied by that history.
Generally, the first state of a norm is the \sname{created} state, which determines whether a norm is relevant to a history. All other states of the norm are conjunctions involving the \sname{created} state.

Listing~\ref{Hercule:consultation} specifies the norms described in Section~\ref{sec:Sidebar}.

\begin{lstlisting}[label={Hercule:consultation},caption={Privacy norm specifications in Hercule, with references to the compact clauses specified above}]
*(\ref{norm:StoreData} \textsf{StoreData})*
commitment StoreData(hospital->patient):
 created: patient.Visit{date}
 detached: physician.Record{patient, item}
 discharged: hospital.Store{item}

*(\ref{norm:DestroyData} \textsf{DestroyData})*
commitment DestroyData(hospital->patient):
 created: hospital.Store{item}
 detached: patient.RequestDeletion{item}
 discharged: hospital.Deleted{item}

*(\ref{norm:Access} \textsf{Access})*
authorization Access(patient->recipient, item):
 created:
  patient.GrantAccess{recipient,item} @ t
 detached:
  recipient.RequestAccess{item} @ t2 > t
  except
  patient.RevokeAccess{recipient,item} @ [t,t2]
 discharged:
  hospital.Shared{item, recipient} @ [t2, t2+10]

*(\ref{norm:Confidentiality} \textsf{Confidentiality})*
prohibition Confidentiality(patient->hospital):
 created: hospital.Store{patient, item}
 violated:
  hospital.Shared{item, recipient}
  except Access(patient->recipient, item):detached
\end{lstlisting}

Listing~\ref{Hercule:consultation} specifies the four norms introduced in Section~\ref{sec:Sidebar}.

\pname{StoreData} is a commitment with three specified states, \sname{created}, \sname{detached}, and \sname{discharged}.
Each state contains an \emph{event} expression used to select matching enactments from the database.
The event expression in the \sname{created} state specifies an event named \ename{Visit}, which is created by \rname{patient} and contains an attribute named \paraname{date}.
Thus, any enactment containing a \ename{Visit} event matching this description is considered as creating an instance of \pname{StoreData}.
Expressions can be composed from simpler ones using operators such as \fsl{and}, \fsl{or}, and \fsl{except}.
The other norm specifications follow the same pattern.

Event expressions can be extended by time expressions, appended with
the \texttt{@} symbol, which either label the time at which an event occurs (as in Access.created, which occurs at time \paraname{t}) or constrain it.  The time expression may perform a simple comparison, as in $t_2 > t_1$, or restrict it to an interval, such as $[t_2, t_2+10]$.

We now explain how an instance of a norm may be created and progress through its states.
An instance of \pname{StoreData} is created when a patient visits the physician---when the patient's agent reports a \ename{Visit} event containing the patient and the date of the visit.
That instance is detached when the physician's agent reports the \ename{Record} event with attributes describing both which patient the record is for, and what item is being recorded.
And, that instance is discharged when the hospital permanently stores the item, as indicated when the hospital's agent reports a \ename{Store} event with the item as an attribute.

Hercule provides special handling of commitments, prohibitions, and authorizations, which all have standard states and semantics, and map well to real-life contracts.
For example, Hercule automatically derives the violated state of a commitment based on the detached and discharged components of its specification.
Hercule could be readily enhanced with additional norm types.

\section{Implementation}
\label{sec:Implementation}
We adopted Hyperledger Fabric as our platform because it provides CouchDB as a document store that supports advanced queries.

Events are recorded by submitting them via transactions to the Hyperledger Fabric network.
Fabric interprets transactions using \emph{chaincode}, plugin programs that execute in containers on every node. The Hercule chaincode adds the submitted events to the appropriate history documents, making changes to the ledger and updating the underlying database to match.
Hyperledger Fabric handles the consensus process by synchronizing and committing changes across all relevant nodes.

CouchDB supports map-reduce processing via \fsl{views} \cite{couchdb}; indexes or collections of derived data that can be queried like a normal collection.
Views are computed by JavaScript functions that are stored in special \fsl{design documents}.
Hercule processes norm specifications to produce a design document for each norm, with one view for each state.
When loaded into the database and queried, each view is applied to the data to produce a separate collection of matching norm instances.
Hercule queries these collections to discover the current states of various norm instances and inform agent behavior, e.g., by detecting past violations by an HCP.

When triggered by a query, CouchDB applies the map and reduce functions of a view to all documents in the database (or incrementally to documents created or updated since the last query) to produce the view collection.
Views are derived from blockchain data but do not modify it, so they do not need to be synchronized across all of the Fabric nodes.
As such, a view needs to be computed not by all nodes in the network, but only those nodes that query it.
The view collections are provided by the underlying CouchDB database independent of the ledger, so agents can query the norm states either by invoking a chaincode query or by directly querying the database.

Listing~\ref{list:StoreData} shows the StoreData part of the design document generated by Hercule for the norm specifications in Listing~\ref{Hercule:consultation}.

\begin{lstlisting}[label={list:StoreData},caption={StoreData design document (generated from Listing~\protect\ref{Hercule:consultation}})]
"StoreData": {
 "language": "javascript",
 "views": {
  ...,
  "violated": {
   "map":
    "function (doc) {
     // created
     doc.Visit
     // detached
     && (doc.Visit && doc.record
     // not discharged
     && !(doc.Visit && doc.Store
          || doc.Visit && doc.Record && doc.Store))
     && emit(doc)
    }"
  }
 }
}
\end{lstlisting}

As a design document, \pname{StoreData} consists of two keys,
\fsf{language} and \fsf{views}.  Each view has the name of the state
as its key, and a single map function implementing the query logic.
Each map function is applied to every document in the database, via
the parameter \fsf{doc}, to produce one or more results via
\fsf{emit}.

Each map function contains a single Boolean expression testing whether
a given \fsf{doc} matches the specified norm state.  For example, the
\fsf{created} function in \pname{StoreData} simply emits all documents
that contain the \pname{Visit} event.  Hercule compiles the various
states following Section~\ref{sec:Norms}.  For example, the
\fsf{detached} function for \pname{StoreData} checks that both
\pname{Visit} and \pname{Record} have occurred.  Similarly, for the
other states.

\section{Conceptual Evaluation}
\label{sec:Conceptual}
We now compare our approach to smart contract systems.

\subsection{Autonomy}
Smart contracts operate automatically, precluding autonomy.
This automation means smart contracts are inviolable, and is sometimes touted as an advantage, but in practice it means that many useful contracts cannot be adequately represented, as the examples in Section~\ref{sec:Smart} show.

Because Hercule does not automatically enforce norms, agents are free to handle violations as they see fit.
Moreover, an enforcement clause may itself be expressed as a norm.
Sharing data without authorization may normally result in punishment, but a hospital may determine that it was necessary for properly responding to a medical emergency.

\subsection{Enactment Scope}
The Bitcoin blockchain implements a single history, in which each transaction is interpreted according to the same rules and can depend on the results of any previous transactions.
Thus, the scope of enactment is broad and could include all the participants and events on the blockchain.

Ethereum includes a similar accounting system, but each Dapp has separate storage \citep{Ethereum}: one Dapp may invoke another but cannot read or modify another's information directly.
The scope of Dapp history could be restricted to selected participants and events or be a long-running open system that every agent may eventually participate in, as in ERC20 tokens, which are themselves full accounting systems \citep{ERC20-Tokens}.

As in Ethereum, Hercule has separate histories for each instance of a compact, which is sufficient for all compacts involving prespecified participants and events.
Thus, the history model reflects the mutual independence of the compacts; if information across histories is essential, the compacts should be specified as one. Of course, an agent participating in two compacts may copy information from one history to another---but that's purely its choice.

\subsection{Operational Model}
In Bitcoin's operational model each transaction depends on the outputs of prior transactions, and must satisfy the scripts of those prior transactions to be validated and added to the ledger \citep{Bitcoin}.

In Ethereum's operational model, a Dapp when invoked may produce whatever new state its code would output, but the length of the computation is bounded by the amount of \emph{gas} provided by the invoker \citep{Ethereum}.
If the computation does not complete before the gas runs out, no updates are recorded but the gas is consumed as fees for the miners.

Hercule does not specify an operational model.
Events must be consistent with the history they are added to, but there are otherwise no constraints on the content of the events themselves or who may submit them.
Norms can be written to identify histories as valid or invalid, but such would only provide warnings to agents (if they looked), not prevent them from making such changes.

Hercule is intended to be customized for specific domains, not operated as a standalone universal system.
A domain-specific deployment could embed an operational model, e.g., requiring that only agents bound to a role in a history can submit events, or using agent signatures to demonstrate consent for changes to the norms.


\section{Performance}
\label{sec:Performance}
Characterizing the performance of a Hyperledger Fabric network is challenging, because of the number of variables and nontrivial interactions between them.
However, unlike smart contract systems, which perform validation and computation when a transaction is submitted prior to the consensus process, Hercule minimally checks each event for consistency during submission, with most of the computation occurring later when the state of a norm is queried.
Furthermore, CouchDB can be used to replicate the data to separate nodes, so that querying can be performed independently of blockchain operations.
Since event submission is trivial in Hercule, and norms can be queried against separate CouchDB instances, Hercule's performance is independent of Hyperledger Fabric's performance.
This motivates our experimental design to measure throughput of norm queries---to verify that Hercule is fast enough to handle the maximum throughput of a Hyperledger Fabric system, and therefore suitable for practical use.

We measured average throughput by creating a new database to prevent caching and
data reuse, loading 200,000 randomly generated enactments, and probing
the \emph{changes per second} statistic while the
database created a view after starting a query.
The number of enactments was selected to achieve a reasonable minimum number of performance samples.
Each enactment was generated with a uniformly distributed degree of
progress---e.g., enactments in which only \ename{Visit} occurred were more common
than those that also included \ename{Shared}.

Our experiment was run on a multiprocess, single-node
CouchDB installation in Docker.  All tests were performed on a
laptop running Gentoo Linux, kernel version 4.19.27, with an Intel
i7-6600U cpu, 16GB of DDR3 memory, and 1TB SSD.
This setup provides a lower bound for performance, since CouchDB can be sharded to run map-reduce operations in parallel across a cluster.

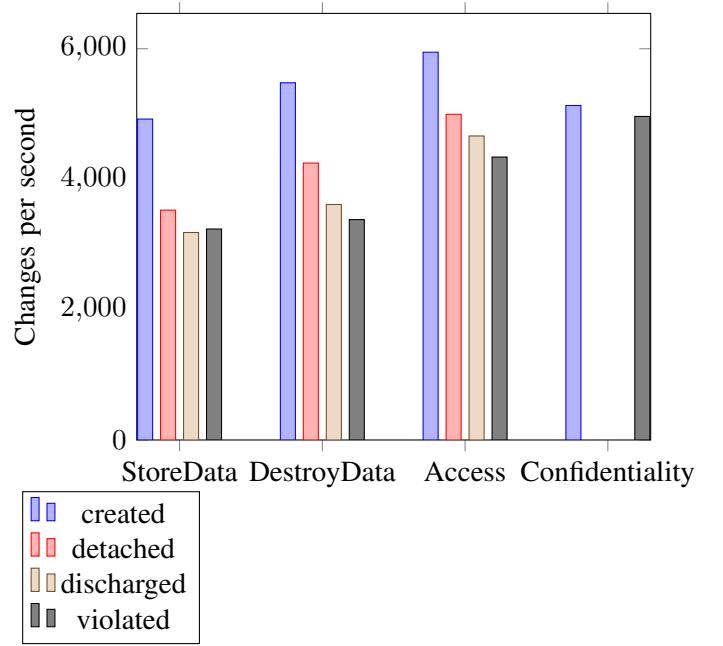
\begin{figure}[htb]
    \centering
\begin{tikzpicture}
\centering
    \begin{axis}[
            width=0.95\linewidth,
            symbolic x coords={StoreData,DestroyData,Access,Confidentiality},
            xtick=data,
            legend style={at={(-0.05,-0.12)},anchor=north},
            ylabel={Changes per second},
            ybar=3pt,
            ymin=0,
            bar width=2mm
        ]
        \addplot coordinates { 
          (StoreData, 4921.66)
          (DestroyData, 5478.32)
          (Access, 5947.52)
          (Confidentiality, 5129.93)
        };
        \addplot coordinates { 
          (StoreData, 3525.61)
          (DestroyData, 4247.24)
          (Access, 4994.67)
        };
        \addplot coordinates { 
          (StoreData, 3182.15)
          (DestroyData, 3611.75)
          (Access, 4662.64)
        };
        \addplot coordinates { 
          (StoreData, 3235.79)
          (DestroyData, 3380.03)
          (Access, 4339.69)
          (Confidentiality, 4962.04)
        };
        \legend{created,detached,discharged,violated}
    \end{axis}
\end{tikzpicture}
    \caption{View construction performance of the example norms.}
    \label{fig:performance}
\end{figure}

Figure~\ref{fig:performance} shows our results grouped by norm and state.
The measured throughput ranges from three to six thousand changes per second across up to eight parallel tasks.
Note that \pname{Confidentiality} does not have detached or discharged states.

Within a given norm later states generally have lower throughput than earlier ones because the later states usually depend on and thus subsume the logic of the earlier ones.
Thus later states require more processing, approximately linear with the size of the logical formula (excepting possible short-circuiting).

Conversely, the later norms have \emph{higher} throughput in a given state than \emph{StoreData} because fewer enactments satisfy their conditions for creation; rejecting objects is the fastest way to process them.

These results correspond to an approximate throughput of three thousand norm state changes per second.
That could be three thousand norms applied to one updated history, or a single norm applied to three thousand changed histories.
Also, if more than one event is added to a history before a norm state is queried again, they will result in only a single change.

A recent performance analysis of Hyperledger Fabric showed that the maximum throughput of a Fabric network is around 400 transactions per second \cite{Nasir2018:fabric-performance}.
Thus, Hercule appears to have sufficient throughput for practical use.
Even if Hercule were slower than Fabric on our test laptop, it scales much better than the Fabric consensus process because each node computes norm states only when queried, and the computation itself is a map-reduce operation that can be distributed across a CouchDB cluster.

\section{Discussion}
\label{sec:discussion}

Hercule demonstrates a possible approach to representing contractual relationships in a way that captures social aspects that cannot be automated, and supports agent autonomy.

A side benefit of Hercule being modeled on norms is that it facilitates incorporating frameworks for intelligent agents that rely on norms to capture social and organizational reasoning capabilities \cite{baldoni2018jacamo+} and compliance monitoring \cite{Modgil+15:monitoring,Dastani+Torroni+Yorke-Smith-18:norms}.
Such approaches typically involve rule-based reasoning about events and often map from norms to cognitive models based on beliefs and goals.

Previous approaches based on logic indicate the viability of a declarative approach.
However, they suffer from the common limitation of adopting the automatic enforcement pattern of smart contracts.
\citet{Governatori2018:declarative-contracts} discuss imperative and declarative smart contracts and their lifecycles, capabilities, and possible implementations, but do not provide a specific model or implementation.
\citet{Purnell2019:declarative-contracts} demonstrate a logic program to implement a will on Ethereum. 
Unlike Hercule, their approach lacks generality and suffers from inefficiency by requiring a logic programming module to be run on all Ethereum nodes to process a transaction.
\Citet{Kruijff2018:commitment-contracts} suggest a commitment-based approach for smart contracts using RuleML as the specification language. They neither support other kinds of norms nor provide an implementation.

As Section~\ref{sec:Histories} explains, Hercule maps each compact to a single history. Some blockchain uses do not readily map to a simple compact. For example, a token accounting system like Bitcoin's is better thought of as a specialized environment with a particular operational model in which unboundedly many transactions can be linked.
One approach to representing such a system would be to treat a transaction as a single history with copies of the input transactions included, and an operational model that prevents the creation of internally inconsistent histories or histories referencing invalid predecessors.
A future direction is to extend the Hercule chaincode interface to enable post-processing on map-reduce queries to handle joins across histories, and produce results agents can use more easily than raw norm states.

Although a typical contract has a lifecycle possibly involving multiple stages of negotiation such as formation, modification, and termination, Hercule focuses on computing the state of active contract instances and leaves the negotiation to external processes.
However, it is straightforward to extend Hercule to support compact modification for a specific application.
For example, the norm specification could be versioned using the blockchain, possibly with an approval process for all parties to consent to a new version.
Then, anyone interested in the compact could evaluate queries according to the most recent version.

Although the design of Hercule supports flexibility in the operational models and enforcement schemes that can be implemented, leaving them out of scope for the core system places additional burdens on the platform implementors and the participating agents.
A future direction is to investigate extensions to compacts supported by Hercule in terms of more sophisticated event syntax and flexible models for
operations and enforcement.

\section{Sources}
All sources are available at \url{https://gitlab.com/masr/hercule}.

\section*{Acknowledgments}
Thanks to NSF (grant IIS-1908374), EPSRC (grant EP/N027965/1), and IBM for support.
Thanks to Aditya Parkhi and Bhavana Balraj for assistance in integrating Hercule with Hyperledger Fabric.


\section*{Author Bios}
\begin{description}
\item [Samuel H.~Christie] is a PhD student at NC State University and
  a Research Associate at the School of Computing and
  Communications at Lancaster University, UK.  Contact him at
  schrist@ncsu.edu.

\item [Amit K.~Chopra] is a Senior Lecturer in the School of Computing
  and Communications at Lancaster University, UK.  Chopra's interests
  include decentralized multiagent systems.  Contact him at amit.chopra@lancaster.ac.uk.

\item [Munindar P.~Singh] is a Professor in Computer Science and a
  co-director of the Science of Security Lablet at NC State
  University.  His research interests include sociotechnical systems.  Singh is a Fellow of AAAI, AAAS, and IEEE, and a former Editor-in-Chief of \emph{IEEE
    Internet Computing} and \emph{ACM Transactions on Internet
    Technology}.  Contact him at singh@ncsu.edu.

\end{description}

\clearpage
\end{document}